\documentclass[preprint,aps]{revtex4}
\usepackage{dcolumn,graphicx}
\def\be {\begin{equation}}
\def\ee {\end{equation}}
\def\mn {{\mu\nu}}
\def\ba {\begin{eqnarray}}
\def\ea {\end{eqnarray}}
\def\cm {{\cal M}}
\def\cl {{\cal L}}
\def\la {\langle}
\def\ra {\rangle}
\def\del {\partial}
\def\eps {\epsilon}
\def\om {\omega}

\def\wt {\widetilde}
\def\oq {\overline{q}^{\,2}}
\def\F {F_\pi}
\def\vq {\vec q}
\def\vk {\vec k}
\def\ep {\epsilon}
\def\de {\delta}
\def\Gm {\Gamma}
\def\sg {\sigma}
\def\omp {\om_\pi}
\def\omh {\om_h}

\begin{document}
\title{Effect of spectral modification of $\rho$ on shear viscosity of a pion gas}
\author{Sukanya Mitra}
\author{Sabyasachi Ghosh}
\author{Sourav Sarkar}
\affiliation{Theoretical Physics Division, Variable Energy Cyclotron Centre, 1/AF Bidhannagar
Kolkata - 700064, India}
\begin{abstract}
We evaluate the shear viscosity of a pion gas in the relativistic kinetic theory
approach. The in-medium propagator of the $\rho$ meson at finite temperature
is used to evaluate the $\pi-\pi$ scattering amplitude in the medium. 
The real and imaginary parts of the self-energy calculated from one-loop diagrams 
are seen to have noticeable effects on the scattering cross-section.
The consequences on temperature dependence of the shear viscosity
evaluated in the Chapman-Enskog and relaxation time approximations are studied. 
\end{abstract}

\maketitle

\section{Introduction}

One of the most exciting revelations that has emerged from experiments at the
Relativistic Heavy Ion Collider (RHIC) is the strongly interacting
nature~\cite{csernai} of the
produced matter. It has properties of an almost perfect fluid characterized by a
rather small value of the shear viscosity over entropy density ratio $\eta/s$, close to
the quantum or KSS bound~\cite{kss}. This interpretation based on the measured
elliptic flow $v_2$ of hadrons in terms of viscous hydrodynamics however
depends sensitively on the value of $\eta/s$.
A lot of interest has been generated
leading to quite a few estimates of the transport coefficients of both
partonic~\cite{arnold1,arnold2} 
as well as hadronic~\cite{prakash,dobado1,dobado2,nakano1,nakano2,itakura,fraile} constituents of strongly interacting matter. In general, the
shear viscosity coefficient is inversely proportional to the cross-section
leading to strongly interacting systems being less viscous than non-interacting
ones, the larger values of the viscous coefficients resulting from the
possibility of momentum transport over larger distances in the latter case.
The interaction cross-section thus plays the pivotal role in the relaxation of
systems towards equilibrium.

Pions form the most important component in a hadronic system. There have been quite a 
few estimates of the viscosity of a pion gas. Almost all these calculations have been
performed in the relativistic kinetic theory approach in which a crucial input is the
$\pi\pi$ differential cross-section.
In~\cite{dobado1,nakano2}, the
scattering amplitude was estimated from the lowest order Lagrangian of chiral
perturbation theory~\cite{weinberg1,gasser} and in~\cite{dobado2} a unitarized amplitude
was employed for a meson gas.
 Phenomenological amplitudes obtained from fits
to phase shift data have been employed in~\cite{itakura} in view of the fact that  
the $\pi\pi$ cross-section estimated from lowest order chiral perturbation theory
is known to
deviate from the experimental data beyond centre of mass energy of 500 MeV
primarily due to the $\rho$ pole which dominates the cross-section in the
energy region between 500-1000 MeV. 

The spectral modification of the $\rho$ in hot and dense matter is known to play the
most important role in explaining the invariant mass spectra of lepton pairs
from relativistic heavy ion collisions in the 
region below the nominal $\rho$ mass~\cite{rapp,ghosh3,sarkar}. 
The lepton pair production in this region is related by vector meson
dominance to two-pion annihilation proceeding through $\rho$-exchange. 
Moreover, medium modification of the $\rho$ spectral function incorporated in
the evaluation of elliptic flow of lepton pairs of low invariant
mass have been recently
shown~\cite{payal} to result in values close to those of hadrons of similar mass,
ascertaining the role of $v_2$ of hadrons in describing the
early stages of heavy ion collisions.
 It is natural to extend this scenario to the evaluation of the transport
coefficients, the temperature dependence of which is
particularly relevant for the electromagnetic probes which are emitted 
at all stages of the collision spanning a range temperatures. 
Investigations with temperature dependent $\pi\pi$ cross-section
involving $\sigma$ exchange have recently been addressed in~\cite{buballa}.
Our intention in this work is to observe the
effect of the in-medium spectral function of the $\rho$ on 
the shear viscosity of a pion gas. 

The first task is to obtain the energy dependent $\pi\pi$ cross-section using a 
phenomenological approach which is close to the experimental value and at the same time
is theoretically amenable to the incorporation of medium effects. 
To this end we consider the scattering to proceed via $\rho$ exchange for which the
invariant amplitude is evaluated using effective interactions. Medium effects
on $\rho$ propagation are introduced through one-loop self-energies to obtain
the modified cross-section at finite temperature. These are discussed in section
II. The corresponding effect on the shear viscosity of a pion gas is 
the subject of section III where it is evaluated using the non-relativistic, 
Chapman-Enskog and relaxation time approximations. We end with a summary in
section IV. Details of the calculation of the $\rho$ self-energy in the medium
has been summarized in the Appendix.

\section{The $\pi\pi$ cross-section with medium effects}

We evaluate the invariant amplitude for $\pi\pi$ scattering using an effective
Lagrangian in which the coupling of the $\rho$ meson to the pions is introduced
through the gauge covariant derivative of the pion field operator to 
obtain~\cite{klingl}
\be
\cl_{\rho\pi\pi}=\frac{ig_\rho}{4}\ Tr[V^\mu,[\del_\mu\Phi,\Phi]] 
\ee
where $Tr$ indicates trace in SU(2) space. The matrix $\Phi$ collects the pion fields
in the form
$\left(\begin{array}{cc}\pi^0&\sqrt{2}\pi^+\\\sqrt{2}
\pi^-&-\pi^0\end{array}\right)$ and $V^\mu$ collects the $\rho$ meson fields
analogously. The coupling constant $g_\rho$=6.05 is fixed from the $\rho\to\pi\pi$
decay width. This interaction leads to $\pi\pi$ scattering diagrams with $\rho$
exchange in the $s$, $t$ and $u$ channels. In these calculations we modify the
$\rho$ propagator $D_\mn^{(0)}=(-g_\mn+q_\mu
q_\nu/m_\rho^2)/(q^2-m_\rho^2+i\eps)$ 
replacing $i\eps$ with $im_\rho\Gamma_\rho(s)$ 
where the two-pion decay width is $\Gamma_\rho(s)=\frac{g_\rho^2}{48\pi
s}(s-4m_\pi^2)^{3/2}$. 
This is done only for $s$-channel $\rho$-exchange diagrams which 
contribute only in the case of total isospin $I=1$.

\begin{figure}
\includegraphics[scale=0.3]{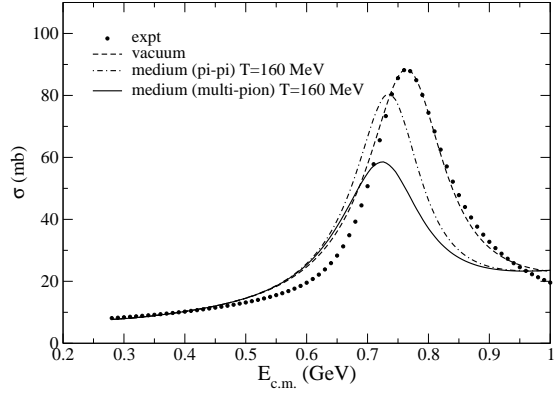}
\caption{The $\pi\pi$ cross-section as a function of centre of mass energy. The
dashed line indicates the cross-section obtained using eq.~\ref{amp} which
agrees well with the experimental values shown by filled circles. The dash-dotted
and solid lines depict the in-medium cross-section for $\pi\pi$ and multi-pion 
loops respectively in the $\rho$ self-energy evaluated at $T$=160 MeV.}
\label{sigmafig}
\end{figure}
 
In order to describe $\pi\pi$ scattering at low energies it is essential to also
include $\sigma$-exchange diagrams. We use the interaction
$\cl_{\sigma\pi\pi}=\frac{1}{2}g_\sigma m_\sigma \vec\pi\cdot\vec\pi \sigma$
with $g_\sigma=2.5$ to calculate the scattering amplitudes where as before we have introduced 
the $\sigma$ width in the  $s$-channel diagram which now appear only for $I=0$.
The values $m_\sigma=450$ MeV and $\Gm_\sigma=550$ MeV that we use are in
conformity with estimates in~\cite{nieves}. 
Going over from the charge 
to the isospin basis it is easy
to write down the matrix elements for all the possible values of the 
total isospin of the two pions. They are,
\ba
\cm_{I=0}&=&2g_\rho^2\left[\frac{s-u}{t-m_\rho^2}+\frac{s-t}{u-m_\rho^2}\right]
+g_\sigma^2 m_\sigma^2\left[\frac{3}{s-m_\sg^2+im_\sg\Gm_\sg}+\frac{1}{t-m_\sg^2}
+\frac{1}{u-m_\sg^2}\right]\nonumber\\
\cm_{I=1}&=&g_\rho^2\left[\frac{2(t-u)}{s-m_\rho^2+im_\rho\Gamma_\rho(s)}+
\frac{t-s}{u-m_\rho^2}-\frac{u-s}{t-m_\rho^2}\right]
+g_\sigma^2 m_\sigma^2\left[\frac{1}{t-m_\sg^2}-\frac{1}{u-m_\sg^2}\right]\nonumber\\
\cm_{I=2}&=&g_\rho^2\left[\frac{u-s}{t-m_\rho^2}+\frac{t-s}{u-m_\rho^2}\right]
+g_\sigma^2 m_\sigma^2\left[\frac{1}{t-m_\sg^2}+\frac{1}{u-m_\sg^2}\right]~.
\label{amp}
\ea
The differential cross-section is then obtained from 
$\frac{d\sigma}{d\Omega}=\overline{|\cm|^2}/64\pi^2 s$ 
where the isospin averaged amplitude is given by
$\overline{|\cm|^2}=\frac{1}{9}\sum(2I+1)\overline{|\cm_I|^2}$.
Ignoring the $I=2$ contribution, the integrated cross-section 
(with an additional factor of 1/2 for identical particles) plotted 
as a function of the centre of mass energy in Fig.~\ref{sigmafig} (dashed line)
is seen to agree 
reasonably well with the one taken from Ref.~\cite{bertsch1}
(shown by filled circles) which corresponds to a resonance saturation
parametrization of isoscalar and isovector phase shifts obtained from various
empirical data involving the $\pi\pi$ system.
We will use this theoretical estimate as a benchmark to study medium effects. 

\begin{figure}
\includegraphics[scale=0.6]{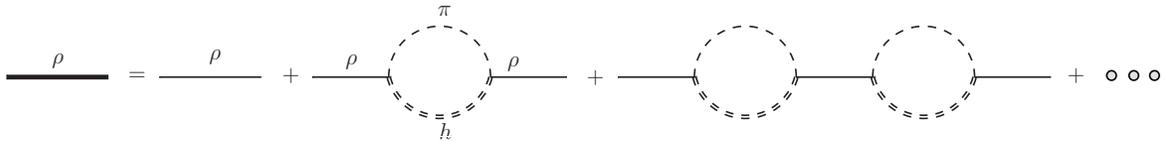}
\caption{The exact $\rho$ propagator with $\pi-h$ loop diagrams for
$h=\pi,\om,h_1,a_1$ mesons.}
\label{dyson}
\end{figure}

The effect of the medium on $\rho$ propagation is quantified through its
self-energy. The standard procedure is to evaluate this quantity by
perturbative methods using effective interactions and then obtain 
 the exact propagator using the Dyson equation, depicted pictorially 
 in Fig.~\ref{dyson}.
We now recollect the main
steps in this scheme omitting details which have been explicitly given in 
\cite{ghosh1,ghosh2}. In the real time
formulation of thermal field theory, all two-point functions assume a $2\times
2$ matrix form~\cite{bellac} which can be diagonalized. The diagonal components
also obey the Dyson equation~\cite{mallik_RT} by means of which
the full propagator $D_\mn$ is obtained as
\be
D_\mn=D^{(0)}_\mn+D^{(0)}_{\mu\sigma}\Pi^{\sigma\lambda}D_{\lambda\nu}
\label{eq:dyson}
\ee
where $D^{(0)}_\mn$ is the vacuum propagator for the $\rho$ meson
and $\Pi^{\sigma\lambda}$ is the self energy function obtained
from one-loop diagrams shown in Fig.~\ref{dyson}.
Following \cite{bellac,ghosh1} we write the in-medium self-energy in terms of
longitudinal and transverse parts
\be
\Pi_\mn=P_\mn \Pi^T+Q_\mn \Pi^L
\ee
where $P_\mn$ and $Q_\mn$ are the transverse and longitudinal projection
tensors respectively. These are defined as~\cite{ghosh1}
\be
P_{\mn}=-g_{\mn}+\frac{q_\mu q_\nu}{q^2}-\frac{q^2}{\oq}\wt u_\mu \wt
u_\nu,~\wt u_\mu=u_\mu-(u\cdot q)q_{\mu}/q^2
\label{defP+Q}
\ee
and
\be
Q_{\mn}=\frac{(q^2)^2}{\oq}\wt u_\mu \wt
u_\nu,~\oq=(u\cdot q)^2-q^2~.
\ee
where $u_\mu$ is four velocity of the thermal bath. It is easy to see that
\be
P_\mn+Q_\mn/q^2=-g_\mn+q_\mu q_\nu/q^2~.
\ee
Note that while $P$ and $Q$ are four-dimensionally transverse, $P$ is also
three-dimensionally transverse while $Q$ is longitudinal. 
Solving (\ref{eq:dyson}), the exact $\rho$ propagator is obtained as
\be
D_\mn(q_0,\vec q)=-\frac{P_\mn}{q^2-m_\rho^2-\Pi^T}-\frac{Q_\mn/q^2}
{q^2-m_\rho^2-q^2\Pi^L}+\frac{q_\mu q_\nu}{m_\rho^2q^2}
\label{exactprop}
\ee
where $\Pi^T$ and $\Pi^L$ can be obtained from the relations
\be
\Pi^T=-\frac{1}{2}(\Pi_\mu^\mu +\frac{q^2}{\bar q^2}\Pi_{00}),~~~~
\Pi^L=\frac{1}{\bar q^2}\Pi_{00} , ~~~\Pi_{00}\equiv u^\mu u^\nu \Pi_{\mn}~.
\label{pitpil}
\ee

As shown in \cite{ghosh1}, the three-momentum dependence of the $\rho$ self-energy is
not substantial for the present case and we can replace $\Pi^T$ and $q^2\Pi^L$ in the above
expression by a self-energy function which is averaged over polarization.
Defining this as
\be
\Pi=\frac{1}{3}(2\Pi^T+q^2\Pi^L)
\ee
and neglecting the non-pole piece in (\ref{exactprop}), the in-medium propagator can be
written as
\be
D_\mn(q_0,\vec q)=\frac{-g_\mn+q_\mu q_\nu/q^2}{q^2-m_\rho^2-{\rm Re}\Pi(q_0,\vec q)+
i{\rm Im}\Pi(q_0,\vec q)}~.
\label{medprop}
\ee
The real part of the self-energy modifies the pole position and the imaginary 
part
embodies the effect of collisions and decay processes by means of which the
$\rho$ is lost or gained in the medium. Using interactions from chiral perturbation theory,
the one-loop self energy was recently calculated~\cite{ghosh1}. 
Some details of the calculation are provided in the Appendix. The imaginary
parts for $\pi\pi$, $\pi\omega$, $\pi h_1$ and $\pi a_1$ loops were obtained from
the discontinuities of the self-energy in the complex energy plane. While for 
the $\pi\pi$ loop,
the contribution at the nominal $\rho$ pole comes from the unitary cut, the
Landau type discontinuity is responsible for contributions from loops with
heavier particles. The mesons $\omega$, $h_1$ and $a_1$ all have negative
G-parity and have substantial $3\pi$ and $\rho\pi$ decay widths~\cite{PDG}. The
self-energies containing these unstable particles in the loop graphs have thus been folded 
with their spectral functions as shown in~\cite{ghosh2}. The contributions from the
loops with heavy mesons may then be considered as a multi-pion contribution to the
$\rho$ self-energy.  
The cross-section obtained by using the in-medium $\rho$-propagator
(\ref{medprop}) in place of the vacuum propagator $D_\mn^{(0)}$ in the 
evaluation of the amplitudes is shown in Fig.~\ref{sigmafig}. We observe a 
small suppression of the peak for the $\pi\pi$ loop and a larger
effect when all the loops (indicated by multi-pion) are considered accompanied by
a small shift in its position. This is due to the temperature dependence 
of the real and
imaginary parts of the self-energy and is manifested as the modified spectral
function of the $\rho$ meson. Extension to the case of finite baryon density
can be made using the results in~\cite{ghosh2}.
Similar reduction of the $\pi\pi$ cross-section for a hot and dense
system were also obtained earlier in~\cite{bertsch2}.

\section{The shear viscosity of a pion gas}

\begin{figure}
\includegraphics[scale=0.3]{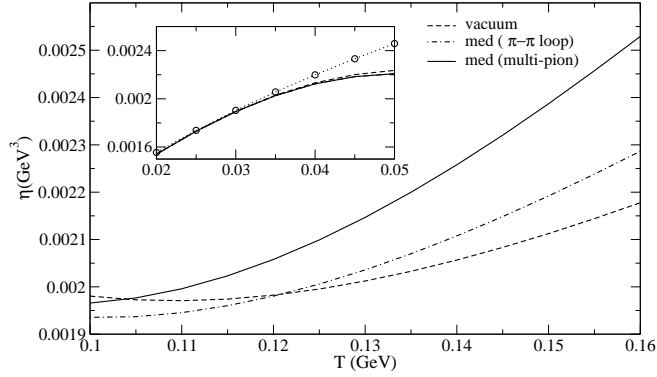}
\caption{The classical estimate of the shear viscosity of a pion gas 
as a function of temperature. The inset shows the behaviour at low temperature.
The dashed line with circles depicts the $\sqrt{T}$ behaviour at low $T$. The
other curves in the inset are the continuations of those shown in the main
figure to lower temperatures.}
\label{eta_nr}
\end{figure}


A first insight into the medium effects on the shear viscosity
may be obtained by employing the
classical estimate of the latter for a dilute gas of pions in terms of the mean
free path $\lambda=1/n\overline{\sigma}$ and is given by
\be
\eta\simeq\frac{1}{3}n\bar p(T)\lambda=\frac{\bar p(T)}{3\overline{\sigma}}
\label{rough_def}
\ee
where the thermally averaged pion momentum and cross-section 
for the process $\pi(p)+\pi(k)\to\pi(p')+\pi(k')$ are given
respectively by
\ba
\bar p(T)&=&\frac{\int d^3p\ |\vec p| f(E_p)}{\int d^3p\ f(E_p)}\nonumber\\
\overline{\sigma}&=&\frac{\int d^3p\ d^3k \ f(E_p)f(E_k)\sigma(s)
(1+f(E_{p'}))(1+f(E_{k'}))}
{\int d^3p\ d^3k f(E_p)f(E_k)(1+f(E_{p'}))(1+f(E_{k'}))} 
\ea
where $f(E)$ denotes the Bose-Einstein distribution and $E_p=\sqrt{\vec p^{\ 2}+m_\pi^2}$
and $E_k=\sqrt{\vec k^{\ 2}+m_\pi^2}$. Manipulation of the exit channel in the 
c.m. frame leads to $E_{p'}=E_{k'}=E_{c.m.}/2=\sqrt{s}/2$. 
We plot the shear viscosity as a function of temperature for a range of values
relevant for the hadronic phase of heavy ion collisions in Fig.~\ref{eta_nr}. 
The behaviour of $\eta$ in this picture is determined by the interplay of the temperature
dependence of $\bar p$ (which essentially behaves as $\sqrt{T}$ since 
$\bar p^{\ 2}/2m\sim kT$ )
and that of $\overline{\sigma}$. 
A substantial effect of the reduced in-medium cross-section 
can be seen in this approximate estimate of $\eta$.
In spite of possible
differences in the averaging procedure, the values of $\eta$ 
for the vacuum case lie within similar range
with those in~\cite{itakura}. The behaviour at low temperatures is shown
separately in the inset of Fig.~\ref{eta_nr}. 
Since the energy of scattering decreases with $T$
as shown in~\cite{itakura}, the cross-section tends to a constant as 
the temperature is lowered (Fig.~\ref{sigmafig}). 
The shear viscosity is then
proportional to $\sqrt{T}$, a situation depicted by the dashed line 
with circles (for arbitrary proportionality constant). 
This $\sqrt{T}$ behaviour of the shear viscosity at low energies has been 
shown analytically using a hard sphere $\pi\pi$ cross-section
in~\cite{dobado1}.
We also note in the inset that all the curves actually converge
as the temperature is lowered showing that medium effects reduce 
with pion density.


In quantum field theory, the shear viscosity is defined
using linear response in terms of the correlation 
function of the (spatial component) of the energy-momentum tensor 
which can be obtained from the Lagrangian density of the 
system~\cite{zubarev,hosoya}. But as
argued in~\cite{jeon,carrington} a perturbative (diagrammatic) evaluation involves a
sum over an infinite set of complicated diagrams in order to obtain the leading
behaviour even in weakly coupled systems. A more efficient approach is to obtain
these by solving the linearized transport equation. 

The relativistic transport equation for 
the phase space distribution $f(x,p)$ of a pion gas is written as~\cite{deGroot}
\be
p^\mu\partial_\mu f(x,p)=C[f]
\label{treq}
\ee
where $C[f]$ is known as the collision term. 
For binary elastic collisions
$p+k\to p'+k'$ the Uehling-Uhlenbeck collision term is given by
\ba
C[f]&=&\int d\Gamma_k\ d\Gamma_{p'}\ d\Gamma_{k'}[f(x,p')f(x,k') \{1+f(x,p)\}
\{1+f(x,k)\}\nonumber\\
&&-f(x,p)f(x,k)\{1+f(x,p')\}\{1+f(x,k')\}]\ W
\ea
where $d\Gamma_q=\frac{d^3q}{(2\pi)^3q_0}$ and the collision rate $W$ is defined as
\[
W=\frac{s}{2}\ \frac{d\sigma}{d\Omega}(2\pi)^6\delta^4(p+k-p'-k')
\]
which includes an explicit factor of $1/2$ for indistinguishable particles.

A system in local equilibrium is described by ideal fluid dynamics and departures
from this state can be dealt with using dissipative dynamics. The
correspondence between non-equilibrium kinetic theory and viscous hydrodynamics
can be studied by considering small departures from equilibrium.
Following~\cite{davesne}, in the first
Chapman-Enskog approximation one takes the non-equilibrium distribution as
\be
f(x,p)=f^{(0)}(x,p)+f^{(0)}(x,p)[1+f^{(0)}(x,p)]\phi(x,p)
\label{ff}
\ee
where $f^{(0)}(x,p)$ is the local equilibrium (Bose) distribution
with space-time dependent parameters and $\phi(x,p)$ is the deviation
therefrom. Putting in (\ref{treq})
the function $\phi(x,p)$ is seen to satisfy
the linearized transport equation 
\be
p^\mu\partial_\mu f^{(0)}(x,p)=-\cl[\phi]
\label{treq2}
\ee
where
\ba
\cl[\phi]=&&f^{(0)}\int d\Gamma_k\ d\Gamma_{p'}\ d\Gamma_{k'}f^{(0)}(x,k)
\{1+f^{(0)}(x,p')\}\{1+f^{(0)}(x,k')\}\nonumber\\
&&[\phi(x,p)+\phi(x,k)-\phi(x,p')-\phi(x,k')]\ W~.
\ea

From general considerations it can be shown that in order to be a solution 
of this equation, the function $\phi$ has to be of the
form $\phi=A\del_\mu u^\mu+C_\mn\la\del^\mu u^\nu\ra$
neglecting terms related to thermal conduction. 
The symbol $\la\cdots\ra$ indicates a space-like symmetrized
and traceless combination. 

To make a connection with the coefficient of viscosity one makes use of the
energy-momentum tensor given by
\be
T^\mn=\int\ d\Gamma_p p^\mu p^\nu f(p)~.
\ee
Putting in the non-equilibrium distribution function (\ref{ff}) into this
equation, $T^\mn$
separates into ideal and dissipative parts. The latter is then compared with 
the most general form~\cite{weinberg2,purnendu} for this quantity
to arrive at the expression
\be
\eta=-\frac{1}{10}\int d\Gamma_p\ C_\mn\la p^\mu p^\nu \ra
f^{(0)}(p)\{1+f^{(0)}(p)\}~.
\ee
Obtaining  $C_\mn$ by solving eq.~(\ref{treq2}) is the most important
component of the calculation. Evaluation of
the advective derivative of the equilibrium distribution function on the left
hand side of this equation results in the integral equation
\be
\cl[C_\mn]=-\la p_\mu p_\nu\ra f^{(0)}(p)\{1+f^{(0)}(p)\}/T
\ee
which can be reduced to algebraic equations by expanding the unknown function
$C_\mn$ in terms of orthogonal polynomials. Similar techniques have been used by many
authors to calculate the transport
coefficients~\cite{dobado1,dobado2,nakano1,nakano2,itakura,buballa}. Using e.g.
generalized Laguerre polynomials (of half-integral order) the
first approximation to the shear viscosity obtained by keeping  the
first term in the series 
is given in the notation of~\cite{polak,davesne} by
\be
\eta=\frac{T}{10}\ \frac{\gamma_0^2}{c_{00}}
\ee
where
\ba
\gamma_0&=&-10\frac{S_3^{-2}(z)}{S_2^{-1}(z)}~~{\rm with}~~z=m_\pi/T~~{\rm and}\nonumber\\
c_{00}&=&I_1(z)+I_2(z)+I_3(z)~.
\ea

\begin{figure}
\includegraphics[scale=0.3]{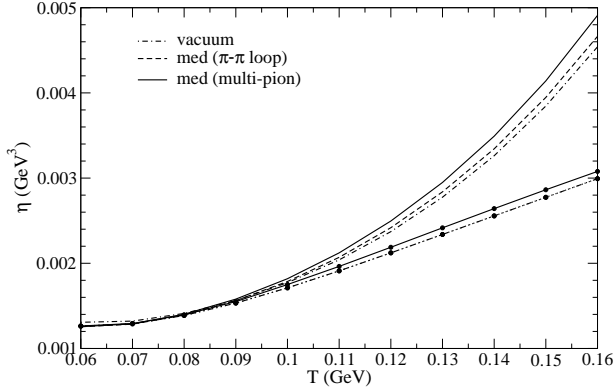}
\caption{The shear viscosity as a function of temperature in the
Chapman-Enskog approximation. The upper set of curves were obtained with
$\psi_{max}\sim 2$ in eq.~\ref{c00_bose} where the dash-dotted line indicates use of the 
vacuum cross-section and the dashed and solid lines 
correspond to in-medium cross-section for the $\pi\pi$ and multi-pion cases 
respectively. The lower set of curves (with circles) were obtained
with $\psi_{max}=\infty$. Here, the dash-dotted line (with circles) 
represents the
vacuum case and the dashed and solid lines (with circles)
correspond to in-medium cross-sections for the $\pi\pi$ and multi-pion cases 
respectively. The difference between the corresponding curves in the upper
and lower sets is a measure of the uncertainty in $\eta$ due to
insufficient information about the cross-section for energies
more than $\sim 1$ GeV.}
\label{CEfig}
\end{figure}

The integrals $I_\alpha(z)$ are given by
\ba
I_\alpha(z)&=&\frac{N_\alpha (z)}{[S_2^{-1}(z)]^2}\int_0^\infty d\psi\ \cosh^3\psi
\sinh^7\psi\int_0^\pi
d\Theta\sin\Theta\frac{1}{2}\frac{d\sigma}{d\Omega}(\psi,\Theta)\int_0^\infty d\chi\ L_\alpha(\chi)\nonumber\\
&&\int_0^{2\pi}
d\phi\int_0^\pi d\theta\sin\theta\frac{e^{2z\cosh\psi\cosh\chi}}
{(e^E-1)(e^F-1)(e^G-1)(e^H-1)}\ M_\alpha(\theta,\Theta)
\label{c00_bose}
\ea
where
\ba
E&=&z(\cosh\psi\cosh\chi-\sinh\psi\sinh\chi\cos\theta)\nonumber\\
F&=&z(\cosh\psi\cosh\chi-\sinh\psi\sinh\chi\cos\theta')\nonumber\\
G&=&E+2z\sinh\psi\sinh\chi\cos\theta\nonumber\\
H&=&F+2z\sinh\psi\sinh\chi\cos\theta'~,
\ea
and the functions $N$, $L$, $M$ and $S$ are defined as
\ba
N_1(z)&=&16z^4,~N_2(z)=16z^4,~N_3(z)=16z^4/3~,\nonumber\\
L_1(\chi)&=&\sinh^2\chi,~L_2(\chi)=\sinh^4\chi,~L_3(\chi)=\sinh^6\chi~,\nonumber\\
M_1(\theta,\Theta)&=&1-\cos^2\Theta,~M_2(\theta,\Theta)=\cos^2\theta+\cos^2\theta'
-2\cos\theta\cos\theta'\cos\Theta~,\nonumber\\
M_3(\theta,\Theta)&=&[\cos^2\theta-\cos^2\theta']^2~~{\rm and}\nonumber\\
S_n^\alpha(z)&=&\sum_{k=1}^\infty k^\alpha K_n(kz)~.
\ea
Here $K_n(x)$ denotes modified Bessel function of order $n$ and
\(
\cos\theta'=\cos\theta\cos\Theta-\sin\theta\sin\Theta\cos\phi~.
\)

We plot $\eta$ versus $T$ in Fig.~\ref{CEfig} obtained in the Chapman-Enskog
approximation showing the effect of the in-medium $\rho$ propagation in
the pion gas. In the upper set of curves the upper limit of integration
over $\psi$ in eq.~(\ref{c00_bose}) is taken to be $\sim 2$ which corresponds to a
center of mass energy ($E_{c.m.}=2m_\pi\cosh\psi$) of $\sim 1$ GeV for $\pi\pi$ scattering up to which we have
obtained a fair description of the cross-section using $\rho$ and $\sigma$
exchange. Here we observe  $\sim 10\%$ change at $T=150$ MeV due to medium effects 
compared to the vacuum when all the loops in the $\rho$ self-energy are considered.
The effect reduces with temperature to less than $5\%$ at 100 MeV and the
curves almost merge with each other as $T$ reaches $\sim 60$ MeV.
The lower set of curves (with circles) in Fig.~\ref{CEfig} depict the corresponding set of 
curves when the actual upper limit of $\psi$ (which is $\infty$) is used. The difference between
the two set of curves indicates the uncertainty in the results on account of 
insufficient information of the cross-section at higher energies.

\begin{figure}
\includegraphics[scale=0.3]{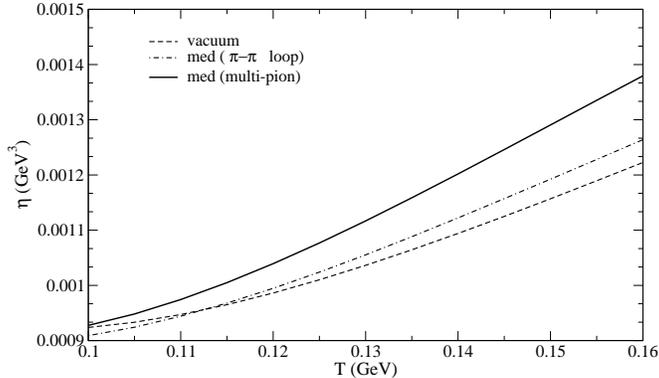}
\caption{The shear viscosity as a function of temperature in the
relaxation time approximation. The dash-dotted and solid lines correspond to the
use of in-medium cross-sections in eq.~\ref{om_relax} for $\pi\pi$ and
multi-pion loops respectively. The dashed line represents the vacuum case.} 
\label{relaxfig}
\end{figure}

We also present results for the
relaxation time approximation which is the simplest way to linearize the transport
equation. Here one assumes that $f(x,p)$ goes over to the equilibrium distribution
$f^{(0)}(x,p)$ as a result of collisions and this takes place over a relaxation
time $\tau(p)$ which is the inverse of the collision frequency $\omega(p)$. The
right hand side of eq.~(\ref{treq}) is then given by
$-E_p\om(p)[f(x,p)-f^{(0)}(x,p)]$. For collisions of the form
$\pi(p)+\pi(k)\to\pi(p')+\pi(k')$ we have
\be
\om(p)=\int\ d\Gamma_k
\frac{\sqrt{s(s-4m_\pi^2)}}{2E_p}f(E_k)(1+f(E_{p'}))(1+f(E_{k'}))\frac{1}{2}\int
d\Omega\frac{d\sigma}{d\Omega}
\label{om_relax}
\ee
and the shear viscosity is given by~\cite{gavin,purnendu,prakash,rupak}
\be
\eta=\frac{1}{15T} \int d\Gamma_p\frac{p^4}{E_p}\tau(p)
f(E_p)(1+f(E_p))~.
\ee
We show the temperature dependence of $\eta$ in the relaxation time
approximation in Fig.~\ref{relaxfig}. The values in this case are lower than
that obtained in the Enskog method though the effect of the medium is
larger. The difference
depends largely on the energy dependence of the cross-section
as noted in~\cite{wiranata}. 

\section{Summary and Outlook}

To summarize, we have used an effective Lagrangian approach to evaluate the
$\pi\pi$ scattering cross-section in which the temperature dependence enters
through the in-medium propagator of the $\rho$ which mediates the interaction
between the pions. This has been used to evaluate the shear viscosity of a
pion gas to the first Chapman-Enskog order. The temperature dependence
with and without medium effects shows a noticeable difference in this approach
as well as in the relaxation time and non-relativistic approaches.

We end by stating that the main focus in this work has been to emphasize the
role of medium modifications of the cross-section in the evaluation of the
transport coefficients. The values of $\eta$ arrived at in the present 
calculation may be modified in a more realistic scenario employing model
independent approaches in the lines of~\cite{dobado3}. In addition, the
treatment has to be extended to include other mesons and
nucleons. Only then can the temperature dependent shear viscosity be included 
in the hydrodynamics instead of constant values used e.g. in the evaluation of
electromagnetic spectra from heavy ion collisions~\cite{sukanya}. Such an effort
is underway and will be reported in due course.

\section{Appendix}
\setcounter{equation}{0}
\renewcommand{\theequation}{A.\arabic{equation}}
Using standard techniques of thermal field theory, the polarization function of the
$\rho$ corresponding to the $\pi-h(=\pi,\om,h_1,a_1)$ loop diagrams shown 
in Fig.~\ref{dyson} is given~\cite{ghosh1,ghosh2} by
\ba
{\Pi}^{\mn}(q_0,\vq)&=&\int\frac{d^3k}{(2\pi)^3}\frac{1}
{4\om_{\pi}\om_{h}}\left[\frac{(1+f(\omp))N^{\mn}_1+f(\omh)N^{\mn}_3}
{q_0 -\om_{\pi}-\om_{h}+i\eta\ep(q_0)}
+\frac{-f(\omp)N^{\mn}_1+f(\omh)N^{\mn}_4}
{q_0-\om_{\pi}+\om_{h}+i\eta\ep(q_0)} 
\right.\nonumber\\
&&+\left. \frac{f(\omp)N^{\mn}_2 -f(\omh)N^{\mn}_3}
{q_0 +\om_{\pi}-\om_{h}+i\eta\ep(q_0)} 
+\frac{-f(\omp)N^{\mn}_2 -(1+f(\omh))N^{\mn}_4}
{q_0 +\om_{\pi}+\om_{h}+i\eta\ep(q_0)}\right]
\label{MM_rho}
\ea
where  $\om_\pi=\sqrt{\vk^2+m_\pi^2}$ and
$\om_h=\sqrt{(\vq-\vk)^2+m_h^2}$ are the Bose distribution
functions. The factor 
$N^{\mn}$ contains tensor structures from the vertices and the vector 
propagators and are given by
\ba
N_{\mn}^{(\pi)}(q,k)&=&\left(\frac{2G_\rho}{m_\rho \F^2}\right)^2
C_{\mn}\nonumber\\
N_{\mn}^{(\om)}(q,k)&=&-4\left(\frac{g_1}{\F}\right)^2(B_{\mn}+q^2k^2
A_{\mn})\nonumber\\
N_{\mn}^{(h_1)}(q,k)&=&-\left(\frac{g_2}{\F}\right)^2(B_{\mn}-\frac{1}{m_{h_1}^2}
C_{\mn})\nonumber\\
N_{\mn}^{(a_1)}(q,k)&=&-2\left(\frac{g_3}{\F}\right)^2(B_{\mn}-\frac{1}{m_{a_1}^2}
C_{\mn})
\ea
where 
\ba
A_{\alpha\beta}(q)&=&-g_{\alpha\beta}+{q_\alpha q_\beta}/{q^2} ,\nonumber\\
B_{\alpha\beta}(q,k)&=&q^2 k_\alpha k_\beta-q\cdot k(q_\alpha k_\beta+k_\alpha q_\beta)
+(q\cdot k)^2g_{\alpha\beta} ,\nonumber\\
C_{\alpha\beta}(q,k)&=&q^4 k_\alpha k_\beta-q^2(q\cdot k)(q_\alpha k_\beta+k_\alpha
q_\beta) +(q\cdot k)^2q_\alpha q_\beta .
\ea
The interaction Lagrangians used to obtain these are given in~\cite{ghosh1}.
The subscript $i(=1,..4)$ on $N^{\mn}$ in (\ref{MM_rho}) correspond to its values for 
$k_0=\om_\pi,-\om_\pi,q_0-\om_h,q_0+\om_h$ respectively. The imaginary part is
easily read off from (\ref{MM_rho}) and is given by
\ba
&& {\rm Im} \Pi^\mn(q_0,\vq)=-\pi\int\frac{d^3 k}{(2\pi)^3 4\om_\pi\om_h}\times 
\nonumber\\
&&[N^\mn_1\{(1-f(\omp)-f(\omh))\de(q_0-\om_\pi-\om_h)
+(f(\omp)-f(\omh))\de(q_0-\om_\pi+\om_h)\}\nonumber\\
&& + N^\mn_2\{(f(\omh)-f(\omp))\de(q_0+\om_\pi-\om_h)
-(1-f(\omp)-f(\omh))\de(q_0+\om_\pi+\om_h)\}]
\label{ImPi_a}
\ea 
where the factors $N^\mn_{3,4}$ have converted to $N^\mn_1$ or $N^\mn_2$ on use
of the associated $\de$-functions. The latter actually define the kinematic
domains where different scattering and decay processes leading to the loss or
gain of the $\rho$ mesons in the medium. The
non-vanishing regions produce branch cuts in the self-energy function. The first
and the fourth terms are non-zero for $q^2>(m_h+m_\pi)^2$ giving rise to the
unitary cut and the second and third terms are non-vanishing for  
$q^2<(m_h-m_\pi)^2$ giving rise to the Landau cut. Whereas the unitary cut is
present in vacuum as well, the Landau cut appears only in the medium. 

Performing the angular integration, the unitary and Landau cut contributions 
for the physically relevant region $q_0,q^2>0$ are
obtained respectively as
\ba
{\rm Im}\Pi^\mn_U (q_0,\vq)&=&-\frac{R^2}{32\pi q^2}\int_{-v}^v dx\ N^\mn(x)
\{1+f(\om)+f(q_0-\om)\}\nonumber\\
&& {\rm for}\ q_0\geq \sqrt{(m_h+m_\pi)^2+|\vq|^2}\nonumber\\
{\rm Im}\Pi^\mn_L (q_0,\vq)&=&\frac{R^2}{32\pi q^2}\int_{-v}^v dx\ N^\mn(x)
\{f(\tilde{\om})-f(q_0+\tilde{\om})\nonumber\\
&& {\rm for}\ |\vq|\leq q_0\leq
\sqrt{(m_h-m_\pi)^2 +|\vq|^2}
\ea
where $x$ is defined in the two cases through $\om=(R^2/2q^2)(q_0+|\vq| x)$
and $\tilde{\om}=(R^2/2q^2)(-q_0 +|\vq| x)$ with $R^2=q^2-m_h^2+m_\pi^2$
and $v=\sqrt{1 -\frac{4q^2 m_\pi ^2}{R^4}}$.
The transverse and longitudinal components for different graphs
can now be obtained using the relations,
\ba
&& A_\mu^\mu =-3,~~~~ A_{00}=\frac{|\vq|^2}{q^2}\\
&& B_\mu^\mu =m_\pi^2 q^2 +\frac{R^4}{2},~~~~ B_{00}=-\frac{|\vq|^2R^4}{4q^2}(1-x^2)\\
&&  C_\mu^\mu =q^2(m_\pi^2 q^2 -\frac{R^4}{4}),~~~~
C_{00}=\frac{|\vq|^2R^4}{4}x^2~.
\ea
The real parts are obtained as principal value integrals which remain 
after removal of the imaginary part from eq.~(\ref{MM_rho}). Note that unlike
the imaginary part, the real part of the self-energy at a given value of $q_0$
receives contribution from all the terms.

To take into account the finite $3\pi$ and $\rho\pi$ decay widths ($\Gm_h$)
of the unstable mesons in the loop, the self-energy is folded with 
their spectral function,
\be
\Pi(q,m_h)= \frac{1}{N_h}\int^{(m_h+2\Gm_h)^2}_{(m_h-2\Gm_h)^2}dM^2\frac{1}{\pi} 
{\rm Im} \left[\frac{1}{M^2-m_h^2 + iM\Gm_h(M) } \right] \Pi(q,M) 
\ee
with $N_h=\displaystyle\int^{(m_h+2\Gm_h)^2}_{(m_h-2\Gm_h)^2}
dM^2\frac{1}{\pi} {\rm Im}\left[\frac{1}{M^2-m_h^2 + iM\Gm_h(M)} \right]$.

\end{document}